\documentstyle[12pt,twoside,fleqn,espcrc1]{article}

\newcommand{\AmS}{{\protect\the\textfont2
 A\kern-.1667em\lower.5ex\hbox{M}\kern-.125emS}}
\newcommand{\newsym}[1]{\ifmmode{#1}\else{$#1$}\fi}
\def\gev{(GeV/$c$)$^2$}
\def\q2{\newsym{Q^2}}
\def\gen{\newsym{G_E^n}}
\def\gmp{\newsym{G_M^p}}
\def\gmn{\newsym{G_M^n}}
\def\gep{\newsym{G_E^p}}
\def\hethree{$^{3}{He}$}

\title{Nucleon Form Factors '99\thanks{Report on the Form Factor session held
during the Nucleon99 workshop, Frascati, Italy, June 1999.}}

\author{Kees de Jager\address{Thomas Jefferson National Accelerator
Facility, Newport News, Va 23606, USA}
        and  Bernard Pire\address{Centre de Physique Th\'eorique, \'Ecole
Polytechnique,
F-91128 Palaiseau, France}\thanks{Unit\'e mixte 7644 du CNRS.}}

\begin{document}

\maketitle

\begin{abstract}
We review recent progress in the experimental knowledge of and theoretical
speculations about nucleon form factors, with special emphasis on the large
$Q^2$ region.

\end{abstract}

%
%

\section{THEORETICAL BACKGROUND}
There is now a long history of continuous progress in the
understanding of electromagnetic form factors at large momentum
transfer. After the pioneering works~\cite{countrule} leading to
the celebrated quark counting rules, the understanding of hard
scattering exclusive processes has been
solidly founded~\cite{pion-ff}. A perturbative QCD subprocess  is
factorized from a wave function-like distribution amplitude
 $\varphi(x_i,Q^{2})$
($x_i$ being the light cone fractions of momentum carried by
valence quarks), the \(Q^{2}\) dependence of which is analysed
in the renormalization group approach. Although an asymptotic
expression emerges from this analysis for the \(x\) dependence of
the distribution, it was quickly understood that the evolution to
the asymptotic \(Q^{2}\) is very slow and that indeed some non
pertubative input is required to get reliable estimates of this
distribution amplitude at measurable \(Q^{2}\).

The severe criticism~\cite{isglle84}
that most of the contributions to the form factor were
coming from end-point regions in the \(x\) integration, especially
when very asymmetric distribution amplitudes were used was answered by Li
and Sterman~\cite{li-ste92}
who proposed  a modified factorization formula
which takes into account Sudakov suppression of elastic scattering
for soft gluon exchange. The resulting formula is, for the pion form factor:
\begin{equation}
F=16\pi C_{F} \int dxdy
\int b_{1}db_{1}\hat{\psi} (x,b_{1})
b_{2}db_{2} \hat{\psi} (y,b_{2}) \hskip 0.265em
\alpha _{S} \hskip 0.265em T_H(b_{1},b_{2},x,y),
\end{equation}
The integration range for the light-cone fractions of momentum $x$ and $y$ goes
from $0$ to $1$. The functions $\hat{\psi} (x,b)$ contain a Sudakov form
factor which
suppresses contributions from large transverse distances $b$. This
improvement leads to
an enlargement of  the domain of applicability of  perturbative QCD
calculations of exclusive processes. Whether accessible data may be
understood within
this formalism, is not yet clear and different strongly motivated conclusions
have been stressed~\cite{Jain}. Let us briefly comment on this.
\begin{itemize}
\item Perturbative corrections are still unknown, and it would not be a great
surprise if they give some enhancement factor; remember the K-factor of the
Drell-Yan
process.
\item The description of transverse size effects through Sudakov factors
and through
intrinsic $k_T$ effects in the wave function may give rise to some double
counting
effects. The phenomenology of Sudakov suppression factors at moderate
transfers is basically
unknown.
\item The Feynman 'soft' process may be a way to rephrase the perturbative
calculation
in some kinematical domain where this latter is not sound. It does however
not seem logical
 to advocate Sudakov suppression of the perturbative process and not
estimate the
corresponding suppression factor in the soft case
\item The concept of nuclear filtering~\cite{jpr} may turn to be very
useful to the
understanding of the free nucleon data. The relative contributions of short
distance
dominated versus soft processes should indeed be differentiated by the
color transparency
 phenomenon. Selecting events where the outgoing hard scattered proton is
not subject to final state
interactions is indeed equivalent to selecting compact configurations which
are characteristic of
the short distance process.

\end{itemize}
In conclusion, it is fair to say that nobody now believes that form factors
are sufficient
to determine the proton distribution amplitude. A comprehensive analysis of
many more data
on different exclusive reactions at large transfers are needed. This in
turn necessitates
high luminosity high duty factor medium energy accelerators~\cite{future}.

\section{TIMELIKE REGION}

The difference between the timelike and
spacelike meson form factors has been analysed~\cite{GP} in  the framework of
perturbative QCD with Sudakov effects included (but only
in the simpler meson case).

In the timelike region, the amplitude for the hard process ruling
\(\gamma^{*}\rightarrow \pi^{+}\pi^{-}\) is simple to deduce from the spacelike
formula:
\begin{equation}
\label{space-like}
T_{H}=16\pi\alpha_{S}C_{F}
{{xQ^{2}}\over{xQ^{2}+{\bf k}^{2}-i\varepsilon}}\hskip 0.265em
{{1}\over{xyQ^{2}+({\bf k}-{\bf l})^{2}-i\varepsilon}},
\end{equation}

\noindent
changing $Q^{2}\rightarrow -W^{2}$. The new feature with respect to the
spacelike form
factor is that the contour of transverse momenta integration now
goes near poles located at either:
\({\bf k}^{2}=xW^{2}+i\varepsilon\) or:
\(({\bf k}-{\bf l})^{2}=xyW^{2}+i\varepsilon\).
Technically, these poles are, except in the end point regions
(\(x,y\rightarrow 0\)), far from the bounds of integration of the
two independent variables \(k={|{\bf k}|}\) and
\(K={|{\bf k}-{\bf l}|}\). Therefore, we may evaluate the integral
by deforming the contour of integration in the complex plane of
each of these variables.

The result of this analysis (for the meson case) is that the asymptotic
behavior is
the same in the timelike and spacelike regions but that the approach to
asymptotia is quite slow and a rather constant enhancement of the
timelike value is expected at measurable large \(Q^{2}\). This study should be
enlarged to the nucleon case where such an enhancement is clearly shown by
experimental data~\cite{Fenice}.

\section{EXPERIMENTAL PROGRESS}
Up until only a few years ago, the quality of available data on
nucleon form factors was quite limited, except for those on the
magnetic form factor of the proton \gmp, which had been accurately
studied~\cite{arnold}
up to over 30 \gev. Accurate measurements of the electric form factor
of the proton \gep\ were restricted to \q2-values below 1 \gev,
because of the \q2-weighting of the contribution from \gmp\ in the
Rosenbluth-separation technique. Studies of
both neutron form factors had to use elastic or quasielastic
scattering off a deuteron, whereby the subtraction of the contribution
from the proton caused sizeable systematic uncertainties in the analysis.

It has been known for quite some time~\cite{polprinc}, that the quality of the
data would be improved significantly by scattering polarized electrons
either from a polarized target or from an unpolarized target while
measuring the polarization of the recoiling or knocked-out nucleon.
However, it has only been in the last few years that polarized beams
became available with high polarization and intensity and the
required polarized targets and recoil polarimeters were developed.

This has resulted in a first batch of new data with high precision.
\gen\ has been measured at low \q2\ at Mainz~\cite{Mainz} and
NIKHEF~\cite{NIKHEF} using
polarized deuteron and \hethree\ targets and neutron polarimeters.
All these experiments used large-acceptance detectors to compensate
for the still limited luminosity, which required extensive Monte Carlo
analysis techniques. Nuclear corrections, which for the deuteron
turned out to be sizeable at very low \q2-values,
amounted~\cite{gloeckle} to $\approx
50\%$ for \hethree\ at \q2 $\approx  0.35$ \gev, but now \gen-data are
available with an accuracy of $\approx 15\%$ up to 0.65 \gev.
\gep\ has been measured in Hall A~\cite{gep} at JLab in a \q2-range up to 3.5
\gev\ using a focal-plane polarimeter to measure the polarization of
the recoiling proton.  The data with a statistical and systematic accuracy
of less than $8\%$ show that \gep\ decreases with \q2\ relative to
\gmp\ and the dipole prediction, indicating that the spatial
distribution of the charge inside the proton extends further than
that of the magnetization. \gmn\ has been accurately measured up to
0.8 \gev\ at Mainz~\cite{gmn1} and Bonn\cite{gmn2} by measuring the ratio
of neutron to proton
knock-out from unpolarized deuterium. However, the two data sets do not
overlap within their error bars and a new measurement of \gmn\ in a
similar \q2-range has recently been performed at JLab~\cite{gmn}, by studying
quasi-elastic scattering of polarized electrons from a polarized
\hethree\ target.

Further experiments have already been scheduled, or can be expected in
a more distant future, to improve the accuracy and/or extend the
\q2-range of the existing data set. \gen\ will be measured at JLab up
to \q2 $\approx 2$ \gev\ in two separate experiments~\cite{genJLab}, one
using a
neutron polarimeter, the other a polarized deuterium target. The
BLAST detector~\cite{blast} at the MIT-Bates facility will provide very
accurate
data in a lower \q2-range, up to $\approx 0.8$ \gev. The JLab \gep\
data set will be extended first to 6 \gev\ with the same
set-up as used in the first experiment~\cite{gepext1}, later
to 10 \gev\ using a lead-glass calorimeter for the detection of the
scattered electron~\cite{gepext2}.

\section{CONCLUSION}
New data on nucleon form factors with an unprecedented precision have (and
will continue to)
become available in an increasing $Q^2$ domain. However, it is still
difficult to
make a precise statement on the applicability of
improved perturbative calculations of the proton form
 factor at available momentum transfers. Future experience, to be gained
 from experiments at JLab at higher energies and at proposed dedicated
 machines~\cite{future}, will provide essential information.
%

%
\end{document}